\ifx\undefined\documentclass
\documentstyle[a4,axodraw]{elsart}
\ifx\undefined\mathrm\newcommand\mathrm[1]{{\rm #1}}\fi
\newcommand\mathcal[1]{{\cal #1}}
\newcommand\lesssim{\stackrel{\lower.7ex\hbox{$<$}}{\lower.7ex\hbox{$\sim$}}}
\else
\documentclass[a4paper]{elsart}
\usepackage{axodraw}
\usepackage{amssymb}
\fi
\newcommand\order{\mathcal{O}}
\newcommand\GeV{\;[\mathrm{GeV}]}

\begin{document}
\thispagestyle{empty}


\begin{flushright}
INLO-PUB-02/96\\
hep-ph/9603411
\end{flushright}

\vspace{2cm}

\begin{frontmatter}
\title{The effect of the Higgs boson\\ on the threshold cross-section\\in 
$W$-pair production}

\author{Wim~Beenakker${}^1$, Geert Jan van Oldenborgh${}^2$}
\address{Instituut-Lorentz, Rijksuniversiteit Leiden, The Netherlands}
\thanks{research supported by a fellowship of the Royal Dutch 
        Academy of Arts and Sciences}
\thanks{research supported by the Stichting FOM}
\begin{abstract}
We investigate the uncertainty induced by the Higgs-boson mass on the 
determination of $m_W$ from the LEP-2 run near the threshold for $W$-pair
production. For a light Higgs boson the Yukawa interaction between the two 
slowly-moving $W$ bosons gives rise to a correction of close to 1\% to the 
total cross-section.  This corresponds to a 15~MeV shift in the deduced $W$ 
mass for a Higgs-boson mass of 60~GeV.
We present a simple approximation for this correction and discuss its validity.
\end{abstract}
\journal{Phys.\ Lett.\ B}
\date{March 1996}
\end{frontmatter}
\clearpage


\section{Introduction}

One of the main experimental results from the LEP-2 collider will be an 
improved measurement of the $W$ mass \cite{LEP2mass}.  Within the Standard 
Model this mass can already be predicted from the existing accurate 
measurements of other precision variables, such as $\alpha$, $G_\mu$, $m_Z$ 
and, through radiative corrections, $m_t$ \cite{LEP1precision}. 
A recent fit gives $m_W = 80.359 \pm 0.051 
{}^{+0.013}_{-0.024}$ GeV, where the first error is dominated by the error in 
$m_t$ and the second one is obtained by varying the Higgs-boson mass in the 
interval $60 < m_H < 1000$ GeV, with central value $m_H = 300$ GeV 
\cite{LEP2mass}.
As can be seen, a significant part of the uncertainty is due to the 
Higgs-boson mass, which also enters through radiative corrections.  One 
expects that an accurate measurement of the $W$-boson mass at LEP-2, combined 
with an improved determination of $m_t$ at Fermilab \cite{CDFtop}, will either 
narrow down the allowed range of $m_H$, or point to physics beyond the 
Standard Model.

In the initial stages of the energy upgrade of the LEP machine, the most 
promising approach seems to be to exploit the sharp rise of the cross-section 
at the threshold for $W$-pair production \cite{LEP2mass}.  One deduces the $W$ 
mass from a comparison of the measured total cross-section with a theoretical 
prediction.  Due to the steep slope, the envisaged statistical error of 5.6\% 
on the total cross-section (for an integrated luminosity of
$200\;\mathrm{pb}^{-1}$) corresponds to a total statistical error on 
$m_W$ of 95 MeV (experimental systematic errors are expected to be much 
smaller).  Of course we would like the theoretical prediction to be much more 
accurate than the projected experimental error.  Moreover, since one has to 
trust this prediction blindly, a thorough analysis of all sources of 
uncertainties is needed.
Effects which certainly contribute at the 1\% level are initial-state 
radiation, non-resonant (`background') diagrams, the Coulomb correction, and 
leading QCD corrections \cite{LEP2WW}.
Initial-state radiation is normally modelled with some kind of 
structure-function or shower algorithm, convoluted with the lowest-order 
matrix element.
The non-resonant contributions have been computed by various groups 
\cite{nonres}.
The Coulomb correction is a simple factor multiplying the matrix element, and 
in the case of $W$-pair production resummation is not even necessary due to 
the finite width of the $W$ boson \cite{Wim&Ansgar&DimaCoulomb}.
Finally, the integrated $\mathcal{O}(\alpha_s)$ corrections are easily 
included.
The as yet unknown initial--final and final--final state interference effects 
should not influence the total cross-section very much \cite{KhozeRad}.

In this letter we investigate another effect, originating from a light Higgs 
boson.  Near threshold such a light particle mediates a sizeable Yukawa 
interaction between the two slowly-moving $W$ bosons. The static potential
associated with this interaction is given by 
\begin{equation}
  V(r) = -\,\frac{m_W^2}{4 \pi v^2 r}\,\mathrm{e}^{- m_H r}\;,
  \label{eq:potential}
\end{equation}
with $m_W^2/(4 \pi v^2) = \alpha/(4\sin^2\theta_w )$.  The mass of the Higgs 
boson determines the range of the interaction, and consequently the size of 
the correction.  For a relatively light Higgs, $m_W \Gamma_W \lesssim m_H^2 
\ll m_W^2$, the correction to the threshold amplitude is larger than the usual 
$\order(\alpha/\pi) = \order(\alpha/4\pi\sin^2\theta_w)$ by a factor $\pi 
m_W/m_H$.  A Higgs boson at the current lower bound, $m_H \approx 60$~GeV, 
would give rise to a correction of the order of 1\%.  Of course LEP-2 itself 
will increase this bound, but at the initial stages at which this threshold 
mass measurement is performed it will not have collected enough luminosity to 
improve on the LEP-1 direct search bound \cite{LEP2Higgs}.

Replacing the Standard-Model Higgs boson by the CP-even neutral Higgs bosons 
of the Minimal Supersymmetric Standard Model (MSSM) does not change the size 
of this correction in a significant way.  If the masses of these two Higgs 
bosons are degenerate one completely recovers the SM expression; if they are 
unequal they each give rise to a similar correction weighted by 
$\cos^2(\alpha-\beta)$ and $\sin^2(\alpha-\beta)$ respectively, with 
$\tan\beta$ the standard ratio of the Higgs vacuum expectation values, and 
$\alpha$ describing the mixing in the CP-even Higgs sector.  
Due to the absence of detectable effects at LEP-1, a light MSSM Higgs-boson 
implies a small coupling to the gauge bosons, and therefore one cannot obtain 
significantly larger effects at LEP-2 than the SM would give.



\section{Approximation}

\begin{figure}
\begin{center}
\begin{picture}(120,100)(0,0)
\put(0,35){\makebox(0,0)[r]{$s = (p_++p_-)^2\;\Biggl\{$}}
\put(40,62){\makebox(0,0)[b]{$\begin{array}{c}
t = (p_+-k_+)^2\\\overbrace{\rule[8pt]{40pt}{0pt}}\end{array}$}}
\put(0,5){\makebox(0,0)[tl]{$p_-$}}
\ArrowLine(0,5)(20,15)
\Vertex(20,15){1}
\ArrowLine(20,15)(20,55)
\put(15,35){\makebox(0,0)[r]{$\nu_e$}}
\Vertex(20,55){1}
\ArrowLine(20,55)(0,65)
\put(0,65){\makebox(0,0)[bl]{$-p_+$}}
\Photon(20,15)(90,15){3}{7}
\put(55,10){\makebox(0,0)[t]{$W$}}
\Photon(20,55)(90,55){3}{7}
\put(55,60){\makebox(0,0)[b]{$W$}}
\Vertex(60,15){1}
\Vertex(60,55){1}
\DashLine(60,15)(60,55){3}
\put(64,35){\makebox(0,0)[l]{$H$}}
\ArrowLine(120,0)(90,15)
\Vertex(90,15){1}
\ArrowLine(90,15)(120,30)
\ArrowLine(120,40)(90,55)
\Vertex(90,55){1}
\ArrowLine(90,55)(120,70)
\put(125,15){\makebox(0,0)[l]{$\biggl\}\;k_-$}}
\put(125,55){\makebox(0,0)[l]{$\biggl\}\;k_+$}}
\end{picture}
\end{center}
\caption{The $t$-channel box diagram that contains the Higgs boson.}
\label{fig:t} 
\end{figure}
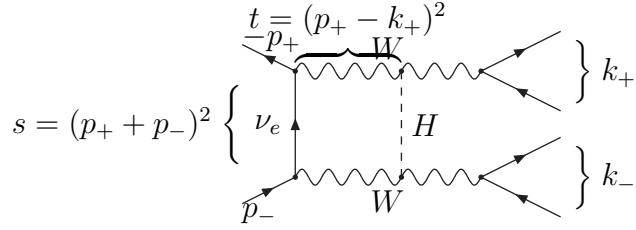

The only radiative corrections to $W$-pair production that depend on the 
Higgs-boson mass are the $W$ and $Z$ self-energies, counterterms dependent on 
these, the $s$-channel vertex corrections depicted in Fig.\ \ref{fig:s}, and 
the $t$-channel box of Fig.\ \ref{fig:t}.

Due to screening, the contributions to the self-energies are at most 
logarithmic in $m_H$ and do not contribute to the light-Higgs enhancement. 
Adopting the LEP-2 input-parameter scenario advocated in Ref.~\cite{LEP2WW},
the $m_H$ dependence is even further reduced. This scenario involves the use
of $\alpha$, $G_\mu$, and $m_Z$ (and the light fermion masses) as input and
treats $m_W$ as free fit parameter. Subsequently, the top-quark mass is
calculated as a function of $m_H$ and $\alpha_s$, using the state-of-the-art 
calculation of $\Delta r$. As such, the logarithmic $m_H$ dependence in the
self-energies has to be largely compensated by the induced variation in 
$m_t$.

The vertex corrections only contribute to $s$-channel invariant 
amplitudes, which are suppressed near threshold by a factor $\beta \equiv 
[(s-s_--s_+)^2 - 4s_+s_-]^{1/2}/s = 
\order([\Delta^2+\Gamma_W^2]^{1/4}/\sqrt{m_W}\,)$, with $s_\pm \equiv k_\pm^2$
the invariant momentum squared of the off-shell $W^\pm$ and 
$\Delta \equiv \sqrt{s}-2 m_W$. 

The dominant contribution will therefore come from the $t$-channel box and, in 
view of the non-relativistic, static nature of the underlying interaction, it 
will be proportional to the dominant lowest-order $t$-channel matrix element%
\footnote{Note that at threshold the SU(2) gauge cancellation 
between $s$- and $t$-channel graphs does not play a role.}, $\mathcal{A}_t$.  
By decomposing 
the amplitude of the $t$-channel box, $\mathcal{A}_{\mathrm{box}}$, into 
standard matrix elements, one can derive the proportionality factor $C_H$. 
Using the notation defined in Fig.\ \ref{fig:t} we find in the Feynman gauge 
($\xi=1$):

\begin{figure}[tb]
\begin{center}
\begin{picture}(130,70)(0,0)
\ArrowLine(0, 5)(30,35)
\Vertex(30,35){1}
\ArrowLine(30,35)(0,65)
\Photon(30,35)(60,35){3}{3.5}
\put(45,30){\makebox(0,0)[t]{$\gamma,Z$}}
\Vertex(60,35){1}
\Photon(60,35)(80,50){3}{3}
\put(80,52){\makebox(0,0)[br]{$W$}}
\Vertex(80,50){1}
\Photon(60,35)(80,20){3}{3}
\put(80,18){\makebox(0,0)[tr]{$W$}}
\Vertex(80,20){1}
\DashLine(80,20)(80,50){3}
\put(82,35){\makebox(0,0)[l]{$H$}}
\Photon(80,20)(100,15){3}{2.5}
\Vertex(100,15){1}
\ArrowLine(130,0)(100,15)
\ArrowLine(100,15)(130,30)
\Photon(80,50)(100,55){3}{2.5}
\Vertex(100,55){1}
\ArrowLine(130,40)(100,55)
\ArrowLine(100,55)(130,70)
\end{picture}
\\
\begin{picture}(130,70)(0,0)
\ArrowLine(0, 5)(30,35)
\Vertex(30,35){1}
\ArrowLine(30,35)(0,65)
\Photon(30,35)(60,35){3}{3.5}
\put(45,30){\makebox(0,0)[t]{$Z$}}
\Vertex(60,35){1}
\DashLine(60,35)(80,50){3}
\put(70,45){\makebox(0,0)[br]{$H$}}
\Vertex(80,50){1}
\Photon(60,35)(80,20){3}{3}
\put(70,23){\makebox(0,0)[tr]{$Z$}}
\Vertex(80,20){1}
\Photon(80,20)(80,50){3}{4}
\put(85,35){\makebox(0,0)[l]{$W$}}
\Photon(80,20)(100,15){3}{2.5}
\Vertex(100,15){1}
\ArrowLine(130,0)(100,15)
\ArrowLine(100,15)(130,30)
\Photon(80,50)(100,55){3}{2.5}
\Vertex(100,55){1}
\ArrowLine(130,40)(100,55)
\ArrowLine(100,55)(130,70)
\end{picture}
\quad
\begin{picture}(130,70)(0,0)
\ArrowLine(0, 5)(30,35)
\Vertex(30,35){1}
\ArrowLine(30,35)(0,65)
\Photon(30,35)(60,35){3}{3.5}
\put(45,30){\makebox(0,0)[t]{$Z$}}
\Vertex(60,35){1}
\Photon(60,35)(80,50){3}{3}
\put(70,48){\makebox(0,0)[br]{$Z$}}
\Vertex(80,50){1}
\DashLine(60,35)(80,20){3}
\put(70,25){\makebox(0,0)[tr]{$H$}}
\Vertex(80,20){1}
\Photon(80,20)(80,50){3}{4}
\put(85,35){\makebox(0,0)[l]{$W$}}
\Photon(80,20)(100,15){3}{2.5}
\Vertex(100,15){1}
\ArrowLine(130,0)(100,15)
\ArrowLine(100,15)(130,30)
\Photon(80,50)(100,55){3}{2.5}
\Vertex(100,55){1}
\ArrowLine(130,40)(100,55)
\ArrowLine(100,55)(130,70)
\end{picture}
\end{center}
\caption[]{The $s$-channel vertex diagrams that contain the Higgs boson. Only 
         the generic diagrams involving $W$ and $Z$ gauge bosons are given.
         By replacing some of the gauge bosons in the loop by the 
         corresponding Higgs ghosts all other diagrams can be obtained.}
\label{fig:s} 
\end{figure}
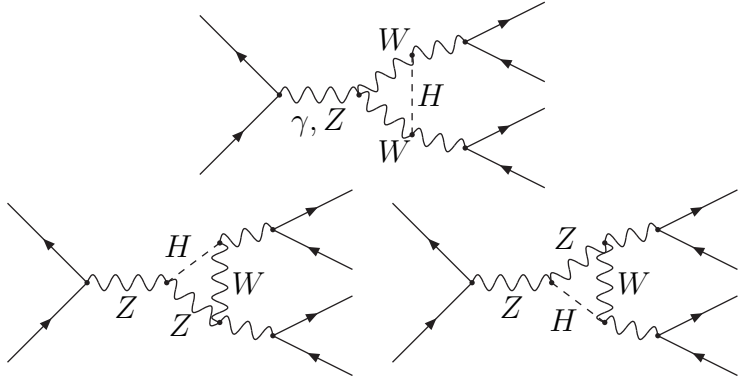

\begin{eqnarray}
\label{eq:full}
    \mathcal{A}_{\mathrm{box}} & \approx & \mathcal{A}_t \,
          \frac{\alpha}{4\pi\sin^2\theta_w} \;
          \frac{- t\,m_W^2/2}{t^2+{s_+}{s_-}-t({s_+}+{s_-}-s)} \times
\nonumber\\&&
\Biggl\{ D_0\bigl[m_H^2(2{s_+}-2t-s) - 2\frac{m_W^2}{s}({s_+}^2-{s_+}{s_-}
            +t{s_-}-t{s_+}) 
\nonumber\\&&\qquad\mbox{}
                 + t(s - 2{s_+} + 2t) + m_W^2(2t+{s_+}+{s_-})\bigr]
\nonumber\\&&\mbox{}
       + C_0^{\nu W^- H}\bigl[3t+{s_-} - 2\frac{t}{s}({s_+}+{s_-}-t) + 
            \frac{2}{s}{s_+}{s_-}\bigr]
\nonumber\\&&\mbox{}
       + C_0^{WWH}   \bigl[s-3{s_+}-{s_-} - 
              \frac{2}{s}(t{s_+}-t{s_-}+{s_+}{s_-}-{s_+}^2)\bigr]
\nonumber\\&&\mbox{}
       + C_0^{W^+\nu H}\bigl[-t+{s_+}-\frac{2}{s}(-2t{s_+}+t^2+{s_+}^2)\bigr]
\nonumber\\&&\mbox{}
       + C_0^{WW\nu} \bigl[-s-2t+2{s_+}\bigr]
          \Biggr\} \equiv \mathcal{A}_t \,C_H\;.
\end{eqnarray}
In this equation, $D_0$ denotes the scalar four-point function corresponding 
to the box diagram, and the $C_0$'s the scalar three-point functions including 
the indicated propagators%
\footnote{Explicit expressions and routines for these functions may be 
found in Ref.\ \cite{C0D0}.}{}.
Superficially, the expression has a pole at the edge of phase space,
$\Delta_3 = -\frac{1}{4}s[t^2+s_+s_--t(s_++s_--s)] = 0$, but the numerator 
also vanishes there to give a finite result.
Near the $W$-pair production threshold the matrix element corresponding to
the $t$-channel box 
\begin{equation}
  \hspace*{-0.8cm}
  \mathcal{A}_{\mathrm{box}} \!\propto\! \int\! \mbox{d}^n l\;
       \frac{(l+p_--k_-)^\mu}{[l^2-m_H^2][(l+k_+)^2-m_W^2][(l-k_-)^2-m_W^2]
       (l+p_--k_-)^2}
\end{equation}
can be simplified considerably by exploiting the fact that to a first 
approximation the two $W$ bosons
are effectively at rest and have an energy close to the beam energy 
[since $|s_{\pm}-m_W^2| \lesssim \order(m_W\Gamma_W)$]. Combined with the 
symmetry
of the integral under the exchange $\,(p_+,k_+) \leftrightarrow (-p_-,-k_-)\,$
this leads to the effective replacement
\begin{equation}
  (l+p_--k_-)^\mu \longrightarrow (p_--k_-)^\mu\,[1+l\cdot(p_--k_-)/t]\;.
\end{equation}
Inserting all the prefactors we arrive at the following threshold 
approximation for $C_H$:
\begin{equation}
\label{eq:approx}
  C_H \approx -\frac{\alpha m_W^2}{8\pi\sin^2\theta_w}\,\left[ (t-m_H^2)D_0
        + C_0^{WWH} - C_0^{WW\nu} \right].
\end{equation}
For a light Higgs ($m_H \ll m_W$) the Yukawa nature of the interaction 
mediated by the Higgs boson dominates if the range of the Yukawa interaction, 
$1/m_H$, is shorter than the characteristic range of a Coulomb-like 
interaction between unstable $W$ bosons, 
$[m_W^2 (\Delta^2+\Gamma_W^2)]^{-1/4}$. If this is the case%
\footnote{If the Higgs were to be significantly lighter, 
$m_H^2 \ll m_W\sqrt{\Delta^2+\Gamma_W^2}$, the Yukawa interaction becomes 
effectively Coulomb-like [see Eq.~(\ref{eq:potential})]. This would 
result in an enhancement of the strength of the $W^+ W^-$ Coulomb interaction 
according to $\alpha \to \alpha + m_W^2/(4 \pi v^2) 
= \alpha + \alpha/(4 \sin^2\theta_w )$.}%
the leading part of Eq.~(\ref{eq:approx}), i.e., the part that 
scales with the range of the interaction, takes on the form 
\begin{displaymath}
  C_H \approx \frac{m_W^2}{4 \pi v^2}\,\frac{m_W}{m_H} 
      = \frac{\alpha}{4 \sin^2\theta_w}\,\frac{m_W}{m_H}\;.
\end{displaymath}


\section{Comparison with the full one-loop result}

In order to check the accuracy of the approximations given in 
Eqs~(\ref{eq:full}) and (\ref{eq:approx}) we would like to compare 
with a full off-shell $\order(\alpha)$ calculation.  
Unfortunately, this calculation is not yet available.  As a first step the
comparison with the full on-shell $\order(\alpha)$ results \cite{BoehmWW} was
performed and excellent agreement was found 
for all production angles of the $W$ bosons and all $\sqrt{s}$ up to 175~GeV 
(better than 0.1\% in $\sigma$, better than 1\% in $\langle \cos\theta_W 
\rangle$).
For the off-shell comparison we restrict ourselves to the factorizable parts 
of the full calculation and calculate these in the (gauge invariant) pole 
scheme \cite{Andre&Geert&Daniel,WWreview}, using the $G_\mu$ renormalization 
scheme.  For a given value of $m_H$ we 
calculate the corresponding corrections and define the full Higgs-boson effect
by subtracting the corrections at $m_H=1$~TeV.  The unknown non-factorizable 
corrections do not depend on the Higgs-boson mass, and will cancel in the 
difference.  The choice of subtraction point, $m_H=1$~TeV, is motivated by 
the fact that heavy Higgs bosons decouple to a large extent in our
calculational scheme (based on the LEP-2 input scenario of 
Ref.~\cite{LEP2WW}). 

Before coming to the discussion of the quality of the approximations, a few
technical remarks are in place. First of all it should be noted that
the strict one-loop pole scheme of Ref.\ \cite{Andre&Geert&Daniel} is not 
well-behaved near the threshold for $W$-pair production. However, a recent
analysis has revealed that the gauge violations of the off-shell tree-level 
amplitude are numerically negligible \cite{LEP2WW,stable,CostasWWV}. As the 
Higgs corrections do not contain any large threshold effects, we have combined 
the off-shell tree-level amplitudes \cite{AndreThesis} with 
the one-loop form factors in the pole scheme.  
In order to make the numerical results more realistic, we improve the 
tree-level results and Higgs-boson effects by inclusion of initial-state 
radiation,  implemented as a 
structure function \cite{KuraevFadinStruct,LeidenEemumu2loop}, and the Coulomb 
correction.  In addition we have included the universal non-resonant graphs, 
i.e., the ones that contribute to all $W$-pair channels.  In the following we 
will refer to these improved tree-level results as `improved Born'.

\begin{table}[tb]
\begin{center}
\begin{tabular}{||c|rrrrr||}
\hline\hline
\multicolumn{6}{||c||}{$m_t\GeV$} \\ \hline\hline
\multicolumn{1}{||c|}{$m_H$}
     & \multicolumn{5}{|c||}{$m_W\GeV$} \\ 
\multicolumn{1}{||c|}{$\hspace*{-6pt}\GeV$} 
     & 80.10 & 80.18 & 80.26 & 80.34 & 80.42 \\
\cline{2-6}
  60 & 119.1 & 133.9 & 148.1 & 161.7 & 174.4 \\
 300 & 137.3 & 151.6 & 165.3 & 178.3 & 190.7 \\
1000 & 154.2 & 168.0 & 181.2 & 193.9 & 206.1 \\
\hline\hline
\end{tabular}
\end{center}
\caption[]{Top-quark mass as a function of the $W$-boson and Higgs-boson 
           masses.}
\label{tab:top}
\end{table}

\begin{table}[tb]
\begin{center}
\begin{tabular}{||c|rrrrr||}
\hline\hline
\multicolumn{6}{||c||}{$\sigma_{\mathrm{tot}}\;[\mathrm{pb}]$} \\ \hline\hline
     & \multicolumn{5}{|c||}{$m_W\GeV$} \\ 
     & 80.10 & 80.18 & 80.26 & 80.34 & 80.42 \\
\hline
\multicolumn{1}{||c|}{$m_H\GeV$} & \multicolumn{5}{|c||}{improved Born} \\
\cline{2-6}
 any & 3.941 & 3.768 & 3.599 & 3.435 & 3.274 \\
\hline
\multicolumn{1}{||c|}{$m_H\GeV$} & \multicolumn{5}{|c||}{Higgs effect:
     approximation Eq. (\ref{eq:approx})} \\
\cline{2-6}
  60 & .0348 & .0335 & .0322 & .0308 & .0293 \\
 300 & .0022 & .0021 & .0021 & .0020 & .0019 \\
1000 & .0002 & .0002 & .0002 & .0002 & .0002 \\
\hline
\multicolumn{1}{||c|}{$m_H\GeV$} & \multicolumn{5}{|c||}{Higgs effect: full 
                                   result} \\
\cline{2-6}
  60 & .0351 & .0343 & .0323 & .0296 & .0265 \\
 300 & .0004 & .0009 & .0010 & .0011 & .0010 \\
1000 &     0 &     0 &     0 &     0 &     0 \\
\hline\hline
\end{tabular}
\end{center}
\caption[]{Higgs effect on the total cross-section at 
           $\protect\sqrt{s}=161$~GeV as a function of the $W$-boson and 
           Higgs-boson masses.}
\label{tab:sigma}
\end{table}

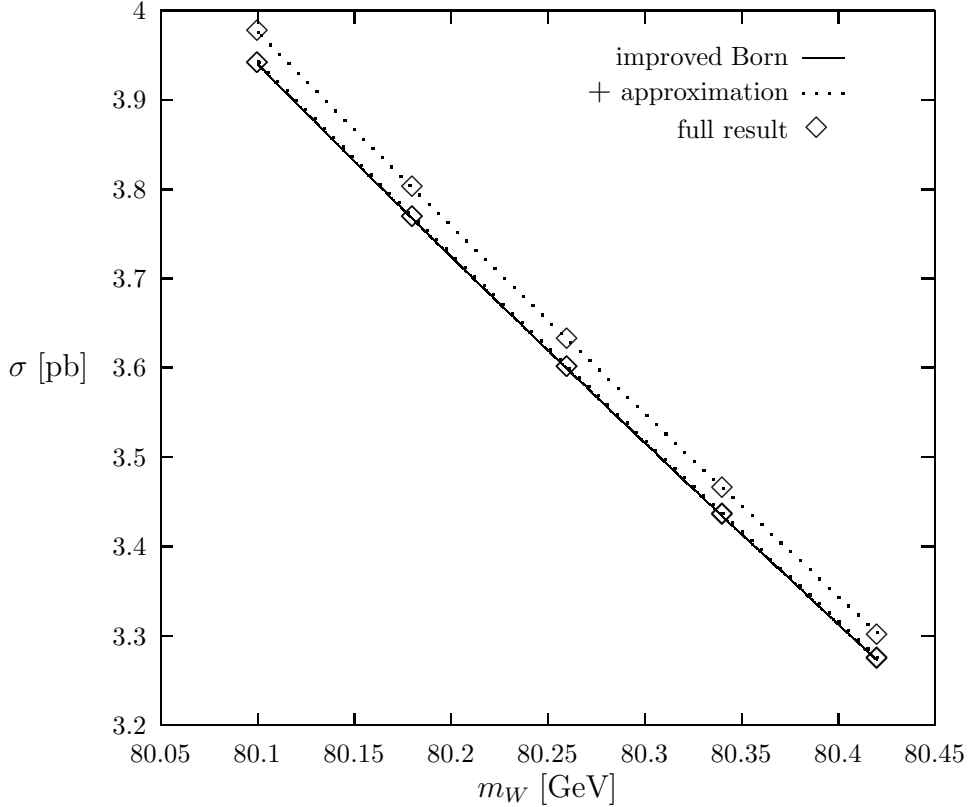
\begin{figure}
\begin{center}
\setlength{\unitlength}{0.240900pt}
\ifx\plotpoint\undefined\newsavebox{\plotpoint}\fi
\sbox{\plotpoint}{\rule[-0.200pt]{0.400pt}{0.400pt}}%
\begin{picture}(1500,1259)(0,0)
\font\gnuplot=cmr10 at 10pt
\gnuplot
\sbox{\plotpoint}{\rule[-0.200pt]{0.400pt}{0.400pt}}%
\put(220.0,113.0){\rule[-0.200pt]{4.818pt}{0.400pt}}
\put(198,113){\makebox(0,0)[r]{3.2}}
\put(1416.0,113.0){\rule[-0.200pt]{4.818pt}{0.400pt}}
\put(220.0,253.0){\rule[-0.200pt]{4.818pt}{0.400pt}}
\put(198,253){\makebox(0,0)[r]{3.3}}
\put(1416.0,253.0){\rule[-0.200pt]{4.818pt}{0.400pt}}
\put(220.0,394.0){\rule[-0.200pt]{4.818pt}{0.400pt}}
\put(198,394){\makebox(0,0)[r]{3.4}}
\put(1416.0,394.0){\rule[-0.200pt]{4.818pt}{0.400pt}}
\put(220.0,534.0){\rule[-0.200pt]{4.818pt}{0.400pt}}
\put(198,534){\makebox(0,0)[r]{3.5}}
\put(1416.0,534.0){\rule[-0.200pt]{4.818pt}{0.400pt}}
\put(220.0,675.0){\rule[-0.200pt]{4.818pt}{0.400pt}}
\put(198,675){\makebox(0,0)[r]{3.6}}
\put(1416.0,675.0){\rule[-0.200pt]{4.818pt}{0.400pt}}
\put(220.0,815.0){\rule[-0.200pt]{4.818pt}{0.400pt}}
\put(198,815){\makebox(0,0)[r]{3.7}}
\put(1416.0,815.0){\rule[-0.200pt]{4.818pt}{0.400pt}}
\put(220.0,955.0){\rule[-0.200pt]{4.818pt}{0.400pt}}
\put(198,955){\makebox(0,0)[r]{3.8}}
\put(1416.0,955.0){\rule[-0.200pt]{4.818pt}{0.400pt}}
\put(220.0,1096.0){\rule[-0.200pt]{4.818pt}{0.400pt}}
\put(198,1096){\makebox(0,0)[r]{3.9}}
\put(1416.0,1096.0){\rule[-0.200pt]{4.818pt}{0.400pt}}
\put(220.0,1236.0){\rule[-0.200pt]{4.818pt}{0.400pt}}
\put(198,1236){\makebox(0,0)[r]{4}}
\put(1416.0,1236.0){\rule[-0.200pt]{4.818pt}{0.400pt}}
\put(220.0,113.0){\rule[-0.200pt]{0.400pt}{4.818pt}}
\put(220,68){\makebox(0,0){80.05}}
\put(220.0,1216.0){\rule[-0.200pt]{0.400pt}{4.818pt}}
\put(372.0,113.0){\rule[-0.200pt]{0.400pt}{4.818pt}}
\put(372,68){\makebox(0,0){80.1}}
\put(372.0,1216.0){\rule[-0.200pt]{0.400pt}{4.818pt}}
\put(524.0,113.0){\rule[-0.200pt]{0.400pt}{4.818pt}}
\put(524,68){\makebox(0,0){80.15}}
\put(524.0,1216.0){\rule[-0.200pt]{0.400pt}{4.818pt}}
\put(676.0,113.0){\rule[-0.200pt]{0.400pt}{4.818pt}}
\put(676,68){\makebox(0,0){80.2}}
\put(676.0,1216.0){\rule[-0.200pt]{0.400pt}{4.818pt}}
\put(828.0,113.0){\rule[-0.200pt]{0.400pt}{4.818pt}}
\put(828,68){\makebox(0,0){80.25}}
\put(828.0,1216.0){\rule[-0.200pt]{0.400pt}{4.818pt}}
\put(980.0,113.0){\rule[-0.200pt]{0.400pt}{4.818pt}}
\put(980,68){\makebox(0,0){80.3}}
\put(980.0,1216.0){\rule[-0.200pt]{0.400pt}{4.818pt}}
\put(1132.0,113.0){\rule[-0.200pt]{0.400pt}{4.818pt}}
\put(1132,68){\makebox(0,0){80.35}}
\put(1132.0,1216.0){\rule[-0.200pt]{0.400pt}{4.818pt}}
\put(1284.0,113.0){\rule[-0.200pt]{0.400pt}{4.818pt}}
\put(1284,68){\makebox(0,0){80.4}}
\put(1284.0,1216.0){\rule[-0.200pt]{0.400pt}{4.818pt}}
\put(1436.0,113.0){\rule[-0.200pt]{0.400pt}{4.818pt}}
\put(1436,68){\makebox(0,0){80.45}}
\put(1436.0,1216.0){\rule[-0.200pt]{0.400pt}{4.818pt}}
\put(220.0,113.0){\rule[-0.200pt]{292.934pt}{0.400pt}}
\put(1436.0,113.0){\rule[-0.200pt]{0.400pt}{270.531pt}}
\put(220.0,1236.0){\rule[-0.200pt]{292.934pt}{0.400pt}}
\put(45,674){\makebox(0,0){$\sigma\;\mathrm{[pb]}$}}
\put(828,12){\makebox(0,0){$m_W\;\mathrm{[GeV]}$}}
\put(220.0,113.0){\rule[-0.200pt]{0.400pt}{270.531pt}}
\put(1206,1161){\makebox(0,0)[r]{improved Born}}
\put(1228.0,1161.0){\rule[-0.200pt]{15.899pt}{0.400pt}}
\put(1345,216){\usebox{\plotpoint}}
\multiput(1342.74,216.59)(-0.553,0.489){15}{\rule{0.544pt}{0.118pt}}
\multiput(1343.87,215.17)(-8.870,9.000){2}{\rule{0.272pt}{0.400pt}}
\multiput(1332.74,225.59)(-0.553,0.489){15}{\rule{0.544pt}{0.118pt}}
\multiput(1333.87,224.17)(-8.870,9.000){2}{\rule{0.272pt}{0.400pt}}
\multiput(1322.92,234.59)(-0.495,0.489){15}{\rule{0.500pt}{0.118pt}}
\multiput(1323.96,233.17)(-7.962,9.000){2}{\rule{0.250pt}{0.400pt}}
\multiput(1313.74,243.59)(-0.553,0.489){15}{\rule{0.544pt}{0.118pt}}
\multiput(1314.87,242.17)(-8.870,9.000){2}{\rule{0.272pt}{0.400pt}}
\multiput(1303.74,252.59)(-0.553,0.489){15}{\rule{0.544pt}{0.118pt}}
\multiput(1304.87,251.17)(-8.870,9.000){2}{\rule{0.272pt}{0.400pt}}
\multiput(1293.74,261.59)(-0.553,0.489){15}{\rule{0.544pt}{0.118pt}}
\multiput(1294.87,260.17)(-8.870,9.000){2}{\rule{0.272pt}{0.400pt}}
\multiput(1283.92,270.59)(-0.495,0.489){15}{\rule{0.500pt}{0.118pt}}
\multiput(1284.96,269.17)(-7.962,9.000){2}{\rule{0.250pt}{0.400pt}}
\multiput(1274.74,279.59)(-0.553,0.489){15}{\rule{0.544pt}{0.118pt}}
\multiput(1275.87,278.17)(-8.870,9.000){2}{\rule{0.272pt}{0.400pt}}
\multiput(1264.74,288.59)(-0.553,0.489){15}{\rule{0.544pt}{0.118pt}}
\multiput(1265.87,287.17)(-8.870,9.000){2}{\rule{0.272pt}{0.400pt}}
\multiput(1254.92,297.59)(-0.495,0.489){15}{\rule{0.500pt}{0.118pt}}
\multiput(1255.96,296.17)(-7.962,9.000){2}{\rule{0.250pt}{0.400pt}}
\multiput(1245.74,306.59)(-0.553,0.489){15}{\rule{0.544pt}{0.118pt}}
\multiput(1246.87,305.17)(-8.870,9.000){2}{\rule{0.272pt}{0.400pt}}
\multiput(1235.92,315.58)(-0.495,0.491){17}{\rule{0.500pt}{0.118pt}}
\multiput(1236.96,314.17)(-8.962,10.000){2}{\rule{0.250pt}{0.400pt}}
\multiput(1225.74,325.59)(-0.553,0.489){15}{\rule{0.544pt}{0.118pt}}
\multiput(1226.87,324.17)(-8.870,9.000){2}{\rule{0.272pt}{0.400pt}}
\multiput(1215.92,334.59)(-0.495,0.489){15}{\rule{0.500pt}{0.118pt}}
\multiput(1216.96,333.17)(-7.962,9.000){2}{\rule{0.250pt}{0.400pt}}
\multiput(1206.74,343.59)(-0.553,0.489){15}{\rule{0.544pt}{0.118pt}}
\multiput(1207.87,342.17)(-8.870,9.000){2}{\rule{0.272pt}{0.400pt}}
\multiput(1196.74,352.59)(-0.553,0.489){15}{\rule{0.544pt}{0.118pt}}
\multiput(1197.87,351.17)(-8.870,9.000){2}{\rule{0.272pt}{0.400pt}}
\multiput(1186.74,361.59)(-0.553,0.489){15}{\rule{0.544pt}{0.118pt}}
\multiput(1187.87,360.17)(-8.870,9.000){2}{\rule{0.272pt}{0.400pt}}
\multiput(1176.92,370.59)(-0.495,0.489){15}{\rule{0.500pt}{0.118pt}}
\multiput(1177.96,369.17)(-7.962,9.000){2}{\rule{0.250pt}{0.400pt}}
\multiput(1167.74,379.59)(-0.553,0.489){15}{\rule{0.544pt}{0.118pt}}
\multiput(1168.87,378.17)(-8.870,9.000){2}{\rule{0.272pt}{0.400pt}}
\multiput(1157.74,388.59)(-0.553,0.489){15}{\rule{0.544pt}{0.118pt}}
\multiput(1158.87,387.17)(-8.870,9.000){2}{\rule{0.272pt}{0.400pt}}
\multiput(1147.92,397.59)(-0.495,0.489){15}{\rule{0.500pt}{0.118pt}}
\multiput(1148.96,396.17)(-7.962,9.000){2}{\rule{0.250pt}{0.400pt}}
\multiput(1138.74,406.59)(-0.553,0.489){15}{\rule{0.544pt}{0.118pt}}
\multiput(1139.87,405.17)(-8.870,9.000){2}{\rule{0.272pt}{0.400pt}}
\multiput(1128.74,415.59)(-0.553,0.489){15}{\rule{0.544pt}{0.118pt}}
\multiput(1129.87,414.17)(-8.870,9.000){2}{\rule{0.272pt}{0.400pt}}
\multiput(1118.74,424.59)(-0.553,0.489){15}{\rule{0.544pt}{0.118pt}}
\multiput(1119.87,423.17)(-8.870,9.000){2}{\rule{0.272pt}{0.400pt}}
\multiput(1108.92,433.59)(-0.495,0.489){15}{\rule{0.500pt}{0.118pt}}
\multiput(1109.96,432.17)(-7.962,9.000){2}{\rule{0.250pt}{0.400pt}}
\multiput(1099.92,442.58)(-0.495,0.491){17}{\rule{0.500pt}{0.118pt}}
\multiput(1100.96,441.17)(-8.962,10.000){2}{\rule{0.250pt}{0.400pt}}
\multiput(1089.74,452.59)(-0.553,0.489){15}{\rule{0.544pt}{0.118pt}}
\multiput(1090.87,451.17)(-8.870,9.000){2}{\rule{0.272pt}{0.400pt}}
\multiput(1079.74,461.59)(-0.553,0.489){15}{\rule{0.544pt}{0.118pt}}
\multiput(1080.87,460.17)(-8.870,9.000){2}{\rule{0.272pt}{0.400pt}}
\multiput(1069.92,470.59)(-0.495,0.489){15}{\rule{0.500pt}{0.118pt}}
\multiput(1070.96,469.17)(-7.962,9.000){2}{\rule{0.250pt}{0.400pt}}
\multiput(1060.74,479.59)(-0.553,0.489){15}{\rule{0.544pt}{0.118pt}}
\multiput(1061.87,478.17)(-8.870,9.000){2}{\rule{0.272pt}{0.400pt}}
\multiput(1050.74,488.59)(-0.553,0.489){15}{\rule{0.544pt}{0.118pt}}
\multiput(1051.87,487.17)(-8.870,9.000){2}{\rule{0.272pt}{0.400pt}}
\multiput(1040.92,497.58)(-0.495,0.491){17}{\rule{0.500pt}{0.118pt}}
\multiput(1041.96,496.17)(-8.962,10.000){2}{\rule{0.250pt}{0.400pt}}
\multiput(1030.92,507.59)(-0.495,0.489){15}{\rule{0.500pt}{0.118pt}}
\multiput(1031.96,506.17)(-7.962,9.000){2}{\rule{0.250pt}{0.400pt}}
\multiput(1021.74,516.59)(-0.553,0.489){15}{\rule{0.544pt}{0.118pt}}
\multiput(1022.87,515.17)(-8.870,9.000){2}{\rule{0.272pt}{0.400pt}}
\multiput(1011.74,525.59)(-0.553,0.489){15}{\rule{0.544pt}{0.118pt}}
\multiput(1012.87,524.17)(-8.870,9.000){2}{\rule{0.272pt}{0.400pt}}
\multiput(1002.93,534.00)(-0.489,0.553){15}{\rule{0.118pt}{0.544pt}}
\multiput(1003.17,534.00)(-9.000,8.870){2}{\rule{0.400pt}{0.272pt}}
\multiput(992.74,544.59)(-0.553,0.489){15}{\rule{0.544pt}{0.118pt}}
\multiput(993.87,543.17)(-8.870,9.000){2}{\rule{0.272pt}{0.400pt}}
\multiput(982.74,553.59)(-0.553,0.489){15}{\rule{0.544pt}{0.118pt}}
\multiput(983.87,552.17)(-8.870,9.000){2}{\rule{0.272pt}{0.400pt}}
\multiput(972.74,562.59)(-0.553,0.489){15}{\rule{0.544pt}{0.118pt}}
\multiput(973.87,561.17)(-8.870,9.000){2}{\rule{0.272pt}{0.400pt}}
\multiput(963.93,571.00)(-0.489,0.553){15}{\rule{0.118pt}{0.544pt}}
\multiput(964.17,571.00)(-9.000,8.870){2}{\rule{0.400pt}{0.272pt}}
\multiput(953.74,581.59)(-0.553,0.489){15}{\rule{0.544pt}{0.118pt}}
\multiput(954.87,580.17)(-8.870,9.000){2}{\rule{0.272pt}{0.400pt}}
\multiput(943.74,590.59)(-0.553,0.489){15}{\rule{0.544pt}{0.118pt}}
\multiput(944.87,589.17)(-8.870,9.000){2}{\rule{0.272pt}{0.400pt}}
\multiput(933.74,599.59)(-0.553,0.489){15}{\rule{0.544pt}{0.118pt}}
\multiput(934.87,598.17)(-8.870,9.000){2}{\rule{0.272pt}{0.400pt}}
\multiput(924.93,608.00)(-0.489,0.553){15}{\rule{0.118pt}{0.544pt}}
\multiput(925.17,608.00)(-9.000,8.870){2}{\rule{0.400pt}{0.272pt}}
\multiput(914.74,618.59)(-0.553,0.489){15}{\rule{0.544pt}{0.118pt}}
\multiput(915.87,617.17)(-8.870,9.000){2}{\rule{0.272pt}{0.400pt}}
\multiput(904.74,627.59)(-0.553,0.489){15}{\rule{0.544pt}{0.118pt}}
\multiput(905.87,626.17)(-8.870,9.000){2}{\rule{0.272pt}{0.400pt}}
\multiput(895.93,636.00)(-0.489,0.553){15}{\rule{0.118pt}{0.544pt}}
\multiput(896.17,636.00)(-9.000,8.870){2}{\rule{0.400pt}{0.272pt}}
\multiput(885.74,646.59)(-0.553,0.489){15}{\rule{0.544pt}{0.118pt}}
\multiput(886.87,645.17)(-8.870,9.000){2}{\rule{0.272pt}{0.400pt}}
\multiput(875.74,655.59)(-0.553,0.489){15}{\rule{0.544pt}{0.118pt}}
\multiput(876.87,654.17)(-8.870,9.000){2}{\rule{0.272pt}{0.400pt}}
\multiput(865.92,664.58)(-0.495,0.491){17}{\rule{0.500pt}{0.118pt}}
\multiput(866.96,663.17)(-8.962,10.000){2}{\rule{0.250pt}{0.400pt}}
\multiput(855.92,674.59)(-0.495,0.489){15}{\rule{0.500pt}{0.118pt}}
\multiput(856.96,673.17)(-7.962,9.000){2}{\rule{0.250pt}{0.400pt}}
\multiput(846.74,683.59)(-0.553,0.489){15}{\rule{0.544pt}{0.118pt}}
\multiput(847.87,682.17)(-8.870,9.000){2}{\rule{0.272pt}{0.400pt}}
\multiput(836.92,692.58)(-0.495,0.491){17}{\rule{0.500pt}{0.118pt}}
\multiput(837.96,691.17)(-8.962,10.000){2}{\rule{0.250pt}{0.400pt}}
\multiput(826.74,702.59)(-0.553,0.489){15}{\rule{0.544pt}{0.118pt}}
\multiput(827.87,701.17)(-8.870,9.000){2}{\rule{0.272pt}{0.400pt}}
\multiput(816.92,711.59)(-0.495,0.489){15}{\rule{0.500pt}{0.118pt}}
\multiput(817.96,710.17)(-7.962,9.000){2}{\rule{0.250pt}{0.400pt}}
\multiput(807.92,720.58)(-0.495,0.491){17}{\rule{0.500pt}{0.118pt}}
\multiput(808.96,719.17)(-8.962,10.000){2}{\rule{0.250pt}{0.400pt}}
\multiput(797.74,730.59)(-0.553,0.489){15}{\rule{0.544pt}{0.118pt}}
\multiput(798.87,729.17)(-8.870,9.000){2}{\rule{0.272pt}{0.400pt}}
\multiput(788.93,739.00)(-0.489,0.553){15}{\rule{0.118pt}{0.544pt}}
\multiput(789.17,739.00)(-9.000,8.870){2}{\rule{0.400pt}{0.272pt}}
\multiput(778.74,749.59)(-0.553,0.489){15}{\rule{0.544pt}{0.118pt}}
\multiput(779.87,748.17)(-8.870,9.000){2}{\rule{0.272pt}{0.400pt}}
\multiput(768.74,758.59)(-0.553,0.489){15}{\rule{0.544pt}{0.118pt}}
\multiput(769.87,757.17)(-8.870,9.000){2}{\rule{0.272pt}{0.400pt}}
\multiput(758.92,767.58)(-0.495,0.491){17}{\rule{0.500pt}{0.118pt}}
\multiput(759.96,766.17)(-8.962,10.000){2}{\rule{0.250pt}{0.400pt}}
\multiput(748.92,777.59)(-0.495,0.489){15}{\rule{0.500pt}{0.118pt}}
\multiput(749.96,776.17)(-7.962,9.000){2}{\rule{0.250pt}{0.400pt}}
\multiput(739.92,786.58)(-0.495,0.491){17}{\rule{0.500pt}{0.118pt}}
\multiput(740.96,785.17)(-8.962,10.000){2}{\rule{0.250pt}{0.400pt}}
\multiput(729.74,796.59)(-0.553,0.489){15}{\rule{0.544pt}{0.118pt}}
\multiput(730.87,795.17)(-8.870,9.000){2}{\rule{0.272pt}{0.400pt}}
\multiput(719.74,805.59)(-0.553,0.489){15}{\rule{0.544pt}{0.118pt}}
\multiput(720.87,804.17)(-8.870,9.000){2}{\rule{0.272pt}{0.400pt}}
\multiput(710.93,814.00)(-0.489,0.553){15}{\rule{0.118pt}{0.544pt}}
\multiput(711.17,814.00)(-9.000,8.870){2}{\rule{0.400pt}{0.272pt}}
\multiput(700.74,824.59)(-0.553,0.489){15}{\rule{0.544pt}{0.118pt}}
\multiput(701.87,823.17)(-8.870,9.000){2}{\rule{0.272pt}{0.400pt}}
\multiput(690.92,833.58)(-0.495,0.491){17}{\rule{0.500pt}{0.118pt}}
\multiput(691.96,832.17)(-8.962,10.000){2}{\rule{0.250pt}{0.400pt}}
\multiput(680.92,843.59)(-0.495,0.489){15}{\rule{0.500pt}{0.118pt}}
\multiput(681.96,842.17)(-7.962,9.000){2}{\rule{0.250pt}{0.400pt}}
\multiput(671.92,852.58)(-0.495,0.491){17}{\rule{0.500pt}{0.118pt}}
\multiput(672.96,851.17)(-8.962,10.000){2}{\rule{0.250pt}{0.400pt}}
\multiput(661.74,862.59)(-0.553,0.489){15}{\rule{0.544pt}{0.118pt}}
\multiput(662.87,861.17)(-8.870,9.000){2}{\rule{0.272pt}{0.400pt}}
\multiput(651.92,871.58)(-0.495,0.491){17}{\rule{0.500pt}{0.118pt}}
\multiput(652.96,870.17)(-8.962,10.000){2}{\rule{0.250pt}{0.400pt}}
\multiput(641.92,881.59)(-0.495,0.489){15}{\rule{0.500pt}{0.118pt}}
\multiput(642.96,880.17)(-7.962,9.000){2}{\rule{0.250pt}{0.400pt}}
\multiput(632.92,890.58)(-0.495,0.491){17}{\rule{0.500pt}{0.118pt}}
\multiput(633.96,889.17)(-8.962,10.000){2}{\rule{0.250pt}{0.400pt}}
\multiput(622.92,900.58)(-0.495,0.491){17}{\rule{0.500pt}{0.118pt}}
\multiput(623.96,899.17)(-8.962,10.000){2}{\rule{0.250pt}{0.400pt}}
\multiput(612.74,910.59)(-0.553,0.489){15}{\rule{0.544pt}{0.118pt}}
\multiput(613.87,909.17)(-8.870,9.000){2}{\rule{0.272pt}{0.400pt}}
\multiput(603.93,919.00)(-0.489,0.553){15}{\rule{0.118pt}{0.544pt}}
\multiput(604.17,919.00)(-9.000,8.870){2}{\rule{0.400pt}{0.272pt}}
\multiput(593.74,929.59)(-0.553,0.489){15}{\rule{0.544pt}{0.118pt}}
\multiput(594.87,928.17)(-8.870,9.000){2}{\rule{0.272pt}{0.400pt}}
\multiput(583.92,938.58)(-0.495,0.491){17}{\rule{0.500pt}{0.118pt}}
\multiput(584.96,937.17)(-8.962,10.000){2}{\rule{0.250pt}{0.400pt}}
\multiput(574.93,948.00)(-0.489,0.553){15}{\rule{0.118pt}{0.544pt}}
\multiput(575.17,948.00)(-9.000,8.870){2}{\rule{0.400pt}{0.272pt}}
\multiput(564.74,958.59)(-0.553,0.489){15}{\rule{0.544pt}{0.118pt}}
\multiput(565.87,957.17)(-8.870,9.000){2}{\rule{0.272pt}{0.400pt}}
\multiput(554.92,967.58)(-0.495,0.491){17}{\rule{0.500pt}{0.118pt}}
\multiput(555.96,966.17)(-8.962,10.000){2}{\rule{0.250pt}{0.400pt}}
\multiput(544.92,977.58)(-0.495,0.491){17}{\rule{0.500pt}{0.118pt}}
\multiput(545.96,976.17)(-8.962,10.000){2}{\rule{0.250pt}{0.400pt}}
\multiput(534.92,987.59)(-0.495,0.489){15}{\rule{0.500pt}{0.118pt}}
\multiput(535.96,986.17)(-7.962,9.000){2}{\rule{0.250pt}{0.400pt}}
\multiput(525.92,996.58)(-0.495,0.491){17}{\rule{0.500pt}{0.118pt}}
\multiput(526.96,995.17)(-8.962,10.000){2}{\rule{0.250pt}{0.400pt}}
\multiput(515.92,1006.58)(-0.495,0.491){17}{\rule{0.500pt}{0.118pt}}
\multiput(516.96,1005.17)(-8.962,10.000){2}{\rule{0.250pt}{0.400pt}}
\multiput(505.92,1016.58)(-0.495,0.491){17}{\rule{0.500pt}{0.118pt}}
\multiput(506.96,1015.17)(-8.962,10.000){2}{\rule{0.250pt}{0.400pt}}
\multiput(495.92,1026.59)(-0.495,0.489){15}{\rule{0.500pt}{0.118pt}}
\multiput(496.96,1025.17)(-7.962,9.000){2}{\rule{0.250pt}{0.400pt}}
\multiput(486.92,1035.58)(-0.495,0.491){17}{\rule{0.500pt}{0.118pt}}
\multiput(487.96,1034.17)(-8.962,10.000){2}{\rule{0.250pt}{0.400pt}}
\multiput(476.92,1045.58)(-0.495,0.491){17}{\rule{0.500pt}{0.118pt}}
\multiput(477.96,1044.17)(-8.962,10.000){2}{\rule{0.250pt}{0.400pt}}
\multiput(467.93,1055.00)(-0.489,0.553){15}{\rule{0.118pt}{0.544pt}}
\multiput(468.17,1055.00)(-9.000,8.870){2}{\rule{0.400pt}{0.272pt}}
\multiput(457.74,1065.59)(-0.553,0.489){15}{\rule{0.544pt}{0.118pt}}
\multiput(458.87,1064.17)(-8.870,9.000){2}{\rule{0.272pt}{0.400pt}}
\multiput(447.92,1074.58)(-0.495,0.491){17}{\rule{0.500pt}{0.118pt}}
\multiput(448.96,1073.17)(-8.962,10.000){2}{\rule{0.250pt}{0.400pt}}
\multiput(437.92,1084.58)(-0.495,0.491){17}{\rule{0.500pt}{0.118pt}}
\multiput(438.96,1083.17)(-8.962,10.000){2}{\rule{0.250pt}{0.400pt}}
\multiput(428.93,1094.00)(-0.489,0.553){15}{\rule{0.118pt}{0.544pt}}
\multiput(429.17,1094.00)(-9.000,8.870){2}{\rule{0.400pt}{0.272pt}}
\multiput(418.74,1104.59)(-0.553,0.489){15}{\rule{0.544pt}{0.118pt}}
\multiput(419.87,1103.17)(-8.870,9.000){2}{\rule{0.272pt}{0.400pt}}
\multiput(408.92,1113.58)(-0.495,0.491){17}{\rule{0.500pt}{0.118pt}}
\multiput(409.96,1112.17)(-8.962,10.000){2}{\rule{0.250pt}{0.400pt}}
\multiput(398.92,1123.58)(-0.495,0.491){17}{\rule{0.500pt}{0.118pt}}
\multiput(399.96,1122.17)(-8.962,10.000){2}{\rule{0.250pt}{0.400pt}}
\multiput(389.93,1133.00)(-0.489,0.553){15}{\rule{0.118pt}{0.544pt}}
\multiput(390.17,1133.00)(-9.000,8.870){2}{\rule{0.400pt}{0.272pt}}
\multiput(379.92,1143.58)(-0.495,0.491){17}{\rule{0.500pt}{0.118pt}}
\multiput(380.96,1142.17)(-8.962,10.000){2}{\rule{0.250pt}{0.400pt}}
\sbox{\plotpoint}{\rule[-0.400pt]{0.800pt}{0.800pt}}%
\put(1206,1106){\makebox(0,0)[r]{$+$ approximation}}
\multiput(1228,1106)(20.756,0.000){4}{\usebox{\plotpoint}}
\put(1345,219){\usebox{\plotpoint}}
\put(1345.00,219.00){\usebox{\plotpoint}}
\put(1329.57,232.88){\usebox{\plotpoint}}
\multiput(1325,237)(-14.676,14.676){0}{\usebox{\plotpoint}}
\put(1314.61,247.25){\usebox{\plotpoint}}
\put(1299.18,261.14){\usebox{\plotpoint}}
\multiput(1296,264)(-15.427,13.885){0}{\usebox{\plotpoint}}
\put(1283.86,275.14){\usebox{\plotpoint}}
\put(1268.78,289.39){\usebox{\plotpoint}}
\multiput(1267,291)(-15.427,13.885){0}{\usebox{\plotpoint}}
\put(1253.53,303.47){\usebox{\plotpoint}}
\put(1238.39,317.65){\usebox{\plotpoint}}
\multiput(1238,318)(-15.427,13.885){0}{\usebox{\plotpoint}}
\put(1222.96,331.53){\usebox{\plotpoint}}
\multiput(1218,336)(-14.676,14.676){0}{\usebox{\plotpoint}}
\put(1208.00,345.90){\usebox{\plotpoint}}
\put(1192.57,359.79){\usebox{\plotpoint}}
\multiput(1189,363)(-15.427,13.885){0}{\usebox{\plotpoint}}
\put(1177.23,373.77){\usebox{\plotpoint}}
\put(1162.55,388.45){\usebox{\plotpoint}}
\multiput(1160,391)(-15.427,13.885){0}{\usebox{\plotpoint}}
\put(1147.39,402.61){\usebox{\plotpoint}}
\put(1132.29,416.84){\usebox{\plotpoint}}
\multiput(1131,418)(-15.427,13.885){0}{\usebox{\plotpoint}}
\put(1116.86,430.72){\usebox{\plotpoint}}
\multiput(1111,436)(-14.676,14.676){0}{\usebox{\plotpoint}}
\put(1101.90,445.09){\usebox{\plotpoint}}
\put(1086.74,459.26){\usebox{\plotpoint}}
\multiput(1082,464)(-15.427,13.885){0}{\usebox{\plotpoint}}
\put(1071.58,473.42){\usebox{\plotpoint}}
\put(1056.59,487.77){\usebox{\plotpoint}}
\multiput(1053,491)(-15.427,13.885){0}{\usebox{\plotpoint}}
\put(1041.16,501.66){\usebox{\plotpoint}}
\put(1026.46,516.27){\usebox{\plotpoint}}
\multiput(1024,519)(-15.427,13.885){0}{\usebox{\plotpoint}}
\put(1011.30,530.43){\usebox{\plotpoint}}
\put(996.27,544.73){\usebox{\plotpoint}}
\multiput(995,546)(-14.676,14.676){0}{\usebox{\plotpoint}}
\put(981.42,559.22){\usebox{\plotpoint}}
\put(965.99,573.11){\usebox{\plotpoint}}
\multiput(965,574)(-14.676,14.676){0}{\usebox{\plotpoint}}
\put(951.27,587.73){\usebox{\plotpoint}}
\put(936.11,601.90){\usebox{\plotpoint}}
\multiput(936,602)(-15.427,13.885){0}{\usebox{\plotpoint}}
\put(921.22,616.32){\usebox{\plotpoint}}
\multiput(917,621)(-15.427,13.885){0}{\usebox{\plotpoint}}
\put(906.26,630.67){\usebox{\plotpoint}}
\put(891.45,645.17){\usebox{\plotpoint}}
\multiput(888,649)(-15.427,13.885){0}{\usebox{\plotpoint}}
\put(876.40,659.44){\usebox{\plotpoint}}
\put(861.32,673.68){\usebox{\plotpoint}}
\multiput(858,677)(-14.676,14.676){0}{\usebox{\plotpoint}}
\put(846.52,688.23){\usebox{\plotpoint}}
\put(831.48,702.52){\usebox{\plotpoint}}
\multiput(829,705)(-15.427,13.885){0}{\usebox{\plotpoint}}
\put(816.31,716.69){\usebox{\plotpoint}}
\put(801.64,731.36){\usebox{\plotpoint}}
\multiput(800,733)(-15.427,13.885){0}{\usebox{\plotpoint}}
\put(786.47,745.53){\usebox{\plotpoint}}
\put(771.80,760.20){\usebox{\plotpoint}}
\multiput(771,761)(-15.427,13.885){0}{\usebox{\plotpoint}}
\put(756.63,774.37){\usebox{\plotpoint}}
\multiput(751,780)(-14.676,14.676){0}{\usebox{\plotpoint}}
\put(741.96,789.04){\usebox{\plotpoint}}
\put(727.04,803.46){\usebox{\plotpoint}}
\multiput(722,808)(-15.427,13.885){0}{\usebox{\plotpoint}}
\put(711.65,817.39){\usebox{\plotpoint}}
\put(697.19,832.23){\usebox{\plotpoint}}
\multiput(693,836)(-14.676,14.676){0}{\usebox{\plotpoint}}
\put(682.31,846.69){\usebox{\plotpoint}}
\put(667.63,861.37){\usebox{\plotpoint}}
\multiput(664,865)(-15.427,13.885){0}{\usebox{\plotpoint}}
\put(652.47,875.53){\usebox{\plotpoint}}
\put(637.79,890.21){\usebox{\plotpoint}}
\multiput(635,893)(-14.676,14.676){0}{\usebox{\plotpoint}}
\put(623.11,904.89){\usebox{\plotpoint}}
\put(608.10,919.21){\usebox{\plotpoint}}
\multiput(605,922)(-13.885,15.427){0}{\usebox{\plotpoint}}
\put(593.67,934.09){\usebox{\plotpoint}}
\put(578.62,948.38){\usebox{\plotpoint}}
\multiput(576,951)(-13.885,15.427){0}{\usebox{\plotpoint}}
\put(564.33,963.40){\usebox{\plotpoint}}
\put(549.30,977.70){\usebox{\plotpoint}}
\multiput(547,980)(-14.676,14.676){0}{\usebox{\plotpoint}}
\put(534.75,992.50){\usebox{\plotpoint}}
\put(520.07,1007.14){\usebox{\plotpoint}}
\multiput(518,1009)(-14.676,14.676){0}{\usebox{\plotpoint}}
\put(505.29,1021.71){\usebox{\plotpoint}}
\put(490.62,1036.38){\usebox{\plotpoint}}
\multiput(489,1038)(-14.676,14.676){0}{\usebox{\plotpoint}}
\put(475.94,1051.06){\usebox{\plotpoint}}
\put(461.68,1066.13){\usebox{\plotpoint}}
\multiput(460,1068)(-15.427,13.885){0}{\usebox{\plotpoint}}
\put(446.62,1080.38){\usebox{\plotpoint}}
\put(431.94,1095.06){\usebox{\plotpoint}}
\multiput(430,1097)(-13.885,15.427){0}{\usebox{\plotpoint}}
\put(417.78,1110.22){\usebox{\plotpoint}}
\put(402.70,1124.47){\usebox{\plotpoint}}
\multiput(401,1126)(-14.676,14.676){0}{\usebox{\plotpoint}}
\put(388.10,1139.22){\usebox{\plotpoint}}
\put(373.77,1154.23){\usebox{\plotpoint}}
\put(372,1156){\usebox{\plotpoint}}
\put(1345,258){\usebox{\plotpoint}}
\put(1345.00,258.00){\usebox{\plotpoint}}
\put(1329.57,271.88){\usebox{\plotpoint}}
\multiput(1325,276)(-14.676,14.676){0}{\usebox{\plotpoint}}
\put(1314.61,286.25){\usebox{\plotpoint}}
\put(1299.18,300.14){\usebox{\plotpoint}}
\multiput(1296,303)(-15.427,13.885){0}{\usebox{\plotpoint}}
\put(1283.86,314.14){\usebox{\plotpoint}}
\put(1268.78,328.39){\usebox{\plotpoint}}
\multiput(1267,330)(-15.427,13.885){0}{\usebox{\plotpoint}}
\put(1253.53,342.47){\usebox{\plotpoint}}
\put(1238.86,357.14){\usebox{\plotpoint}}
\multiput(1238,358)(-15.427,13.885){0}{\usebox{\plotpoint}}
\put(1223.47,371.07){\usebox{\plotpoint}}
\multiput(1218,376)(-14.676,14.676){0}{\usebox{\plotpoint}}
\put(1208.51,385.44){\usebox{\plotpoint}}
\put(1193.08,399.33){\usebox{\plotpoint}}
\multiput(1189,403)(-15.427,13.885){0}{\usebox{\plotpoint}}
\put(1177.72,413.28){\usebox{\plotpoint}}
\put(1163.04,427.96){\usebox{\plotpoint}}
\multiput(1160,431)(-15.427,13.885){0}{\usebox{\plotpoint}}
\put(1147.88,442.12){\usebox{\plotpoint}}
\put(1132.80,456.38){\usebox{\plotpoint}}
\multiput(1131,458)(-15.427,13.885){0}{\usebox{\plotpoint}}
\put(1117.38,470.26){\usebox{\plotpoint}}
\put(1102.85,485.05){\usebox{\plotpoint}}
\multiput(1102,486)(-15.427,13.885){0}{\usebox{\plotpoint}}
\put(1087.52,499.03){\usebox{\plotpoint}}
\put(1072.09,512.92){\usebox{\plotpoint}}
\multiput(1072,513)(-13.885,15.427){0}{\usebox{\plotpoint}}
\put(1057.67,527.80){\usebox{\plotpoint}}
\multiput(1053,532)(-15.427,13.885){0}{\usebox{\plotpoint}}
\put(1042.24,541.69){\usebox{\plotpoint}}
\put(1027.43,556.19){\usebox{\plotpoint}}
\multiput(1024,560)(-15.427,13.885){0}{\usebox{\plotpoint}}
\put(1012.38,570.46){\usebox{\plotpoint}}
\put(997.66,585.04){\usebox{\plotpoint}}
\multiput(995,588)(-15.427,13.885){0}{\usebox{\plotpoint}}
\put(982.53,599.22){\usebox{\plotpoint}}
\put(967.49,613.51){\usebox{\plotpoint}}
\multiput(965,616)(-14.676,14.676){0}{\usebox{\plotpoint}}
\put(952.65,628.02){\usebox{\plotpoint}}
\put(937.65,642.35){\usebox{\plotpoint}}
\multiput(936,644)(-15.427,13.885){0}{\usebox{\plotpoint}}
\put(922.48,656.52){\usebox{\plotpoint}}
\put(907.81,671.19){\usebox{\plotpoint}}
\multiput(907,672)(-15.427,13.885){0}{\usebox{\plotpoint}}
\put(892.88,685.58){\usebox{\plotpoint}}
\multiput(888,691)(-15.427,13.885){0}{\usebox{\plotpoint}}
\put(877.99,700.01){\usebox{\plotpoint}}
\put(862.83,714.17){\usebox{\plotpoint}}
\multiput(858,719)(-14.676,14.676){0}{\usebox{\plotpoint}}
\put(848.15,728.85){\usebox{\plotpoint}}
\put(833.19,743.22){\usebox{\plotpoint}}
\multiput(829,747)(-14.676,14.676){0}{\usebox{\plotpoint}}
\put(818.31,757.69){\usebox{\plotpoint}}
\put(803.31,772.02){\usebox{\plotpoint}}
\multiput(800,775)(-14.676,14.676){0}{\usebox{\plotpoint}}
\put(788.47,786.53){\usebox{\plotpoint}}
\put(773.80,801.20){\usebox{\plotpoint}}
\multiput(771,804)(-15.427,13.885){0}{\usebox{\plotpoint}}
\put(758.63,815.37){\usebox{\plotpoint}}
\put(743.96,830.04){\usebox{\plotpoint}}
\multiput(742,832)(-14.676,14.676){0}{\usebox{\plotpoint}}
\put(729.14,844.57){\usebox{\plotpoint}}
\put(714.12,858.88){\usebox{\plotpoint}}
\multiput(712,861)(-14.676,14.676){0}{\usebox{\plotpoint}}
\put(699.44,873.56){\usebox{\plotpoint}}
\put(684.34,887.79){\usebox{\plotpoint}}
\multiput(683,889)(-13.885,15.427){0}{\usebox{\plotpoint}}
\put(670.12,902.88){\usebox{\plotpoint}}
\put(655.00,917.10){\usebox{\plotpoint}}
\multiput(654,918)(-14.676,14.676){0}{\usebox{\plotpoint}}
\put(640.28,931.72){\usebox{\plotpoint}}
\put(625.60,946.40){\usebox{\plotpoint}}
\multiput(625,947)(-14.676,14.676){0}{\usebox{\plotpoint}}
\put(610.72,960.86){\usebox{\plotpoint}}
\put(596.26,975.71){\usebox{\plotpoint}}
\multiput(596,976)(-14.676,14.676){0}{\usebox{\plotpoint}}
\put(581.37,990.16){\usebox{\plotpoint}}
\multiput(576,995)(-13.885,15.427){0}{\usebox{\plotpoint}}
\put(566.95,1005.05){\usebox{\plotpoint}}
\put(552.27,1019.73){\usebox{\plotpoint}}
\put(537.11,1033.90){\usebox{\plotpoint}}
\multiput(537,1034)(-13.885,15.427){0}{\usebox{\plotpoint}}
\put(522.94,1049.06){\usebox{\plotpoint}}
\put(508.27,1063.73){\usebox{\plotpoint}}
\multiput(508,1064)(-14.676,14.676){0}{\usebox{\plotpoint}}
\put(493.59,1078.41){\usebox{\plotpoint}}
\multiput(489,1083)(-14.676,14.676){0}{\usebox{\plotpoint}}
\put(478.92,1093.08){\usebox{\plotpoint}}
\put(464.50,1108.00){\usebox{\plotpoint}}
\put(450.08,1122.92){\usebox{\plotpoint}}
\multiput(450,1123)(-14.676,14.676){0}{\usebox{\plotpoint}}
\put(435.16,1137.35){\usebox{\plotpoint}}
\multiput(430,1142)(-13.885,15.427){0}{\usebox{\plotpoint}}
\put(420.75,1152.25){\usebox{\plotpoint}}
\put(406.07,1166.93){\usebox{\plotpoint}}
\put(391.40,1181.60){\usebox{\plotpoint}}
\multiput(391,1182)(-13.885,15.427){0}{\usebox{\plotpoint}}
\put(376.99,1196.51){\usebox{\plotpoint}}
\put(372,1201){\usebox{\plotpoint}}
\sbox{\plotpoint}{\rule[-0.600pt]{1.200pt}{1.200pt}}%
\put(1206,1051){\makebox(0,0)[r]{full result}}
\put(1250,1051){\raisebox{-.8pt}{\makebox(0,0){$\Diamond$}}}
\put(372,1153){\raisebox{-.8pt}{\makebox(0,0){$\Diamond$}}}
\put(615,910){\raisebox{-.8pt}{\makebox(0,0){$\Diamond$}}}
\put(858,674){\raisebox{-.8pt}{\makebox(0,0){$\Diamond$}}}
\put(1102,442){\raisebox{-.8pt}{\makebox(0,0){$\Diamond$}}}
\put(1345,216){\raisebox{-.8pt}{\makebox(0,0){$\Diamond$}}}
\put(372,1153){\raisebox{-.8pt}{\makebox(0,0){$\Diamond$}}}
\put(615,911){\raisebox{-.8pt}{\makebox(0,0){$\Diamond$}}}
\put(858,675){\raisebox{-.8pt}{\makebox(0,0){$\Diamond$}}}
\put(1102,444){\raisebox{-.8pt}{\makebox(0,0){$\Diamond$}}}
\put(1345,218){\raisebox{-.8pt}{\makebox(0,0){$\Diamond$}}}
\put(372,1202){\raisebox{-.8pt}{\makebox(0,0){$\Diamond$}}}
\put(615,958){\raisebox{-.8pt}{\makebox(0,0){$\Diamond$}}}
\put(858,719){\raisebox{-.8pt}{\makebox(0,0){$\Diamond$}}}
\put(1102,484){\raisebox{-.8pt}{\makebox(0,0){$\Diamond$}}}
\put(1345,254){\raisebox{-.8pt}{\makebox(0,0){$\Diamond$}}}
\end{picture}
\end{center}
\caption[]{Total cross-section at $\protect\sqrt{s}=161$~GeV as a function of 
           the $W$-boson mass for a Higgs-boson mass of 60 and 300~GeV; 
           the latter case yields results that are virtually 
           indistinguishable from the improved-Born curve.}
\label{fig:sigma} 
\end{figure}

Using for the input parameters the values specified in Ref.~\cite{LEP2WW}, we 
checked the effect of the different approximations at $\sqrt{s}=161$~GeV
for a sample of Monte Carlo points and for the total cross-section. The 
agreement between the full expression for $C_H$, Eq.~(\ref{eq:full}), and the 
approximation, Eq.~(\ref{eq:approx}), is excellent for $\sigma_{\mathrm{tot}}$ 
($\ll0.1\%$), and better than about 0.15\% for almost all of phase space.  
Adding the box amplitudes not proportional to the tree-level $t$-channel graph 
does not significantly change the agreement. However, the other 
$m_H$-dependent graphs are seen to be more important than in the on-shell 
approximation.  Overall, the simple approximation reproduces the full result 
to about 0.1\% in $\sigma_{\mathrm{tot}}$, but only 0.4\% 
for individual points in phase space.

For the approximation to hold it is essential that, for a given $m_W$, the 
top-quark mass is varied along with the Higgs-boson mass to nullify any 
deviations from existing precision data.  The 
values used are given in Table \ref{tab:top} \cite{LEP2WW}.  
We note that some values of the top-quark mass are clearly incompatible with 
the direct measurement \cite{CDFtop}, but for any value of $m_H$ there is a 
range of allowed values of $m_W$. For a light Higgs, for instance, there is a 
preference for the higher $m_W$ values, which is supported by the present
CDF data \cite{CDFMW95}.

{}From Table \ref{tab:sigma} and Figure \ref{fig:sigma} one can see that the 
effect of a light Higgs boson ($m_H=60$~GeV) is to increase the total 
cross-section at $\sqrt{s}=161$~GeV by about 0.9\%.  This correction rapidly 
diminishes for increasing Higgs-boson mass, for $m_H=300$ GeV it is negligible 
($<0.1$\%).  All this can be translated into an uncertainty on the 
determination of $m_W$ from the LEP-2 threshold run.  A measured total 
cross-section will correspond to
\begin{equation}
	m_W\,(m_H = 300\;\mathrm{GeV})
		_{\textstyle -0\;\mathrm{MeV}\, (m_H = 1000\;\mathrm{GeV})}
		^{\textstyle +15\;\mathrm{MeV}\, (m_H = 60\;\mathrm{GeV})}
\;.
\end{equation}
After the higher-energy LEP-2 runs have taken place, the improved knowledge of 
$m_H$ can be used for an {\it a posteriori\/} reduction of the $m_H$-dependece 
of the threshold measurement.  An increase of the lower bound to $m_H>90$ GeV, 
for instance, would reduce the uncertainty by roughly a factor of two.


\section{Conclusion}

We have investigated one of the possible effects that could influence the 
total cross-section at the $W$-pair threshold at the 1\% level: the 
corrections due to a light Higgs boson.  These corrections are in fact 
slightly smaller (0.8--0.9\% depending on $m_W$).  The dominant effect 
comes from the $t$-channel box.  This effect can be modelled quite easily with 
the simple approximation Eq.~(\ref{eq:approx}), which is accurate to better 
than 0.1\% in the total cross-section (0.4\% in some regions of phase space).  
This simple correction term is available as part of the generator WWF 2.3 
\cite{WWFpickup}.


{
\frenchspacing
 \newcommand{\xxx}[1]{{\it(#1)\/}}
 \newcommand{\zp}[3]{{\sl Z. Phys.} {\bf #1} (19#2) #3}
 \newcommand{\np}[3]{{\sl Nucl. Phys.} {\bf #1} (19#2)~#3}
 \newcommand{\pl}[3]{{\sl Phys. Lett.} {\bf #1} (19#2) #3}
 \newcommand{\pr}[3]{{\sl Phys. Rev.} {\bf #1} (19#2) #3}
 \newcommand{\prl}[3]{{\sl Phys. Rev. Lett.} {\bf #1} (19#2) #3}
 \newcommand{\fp}[3]{{\sl Fortschr. Phys.} {\bf #1} (19#2) #3}
 \newcommand{\nc}[3]{{\sl Nuovo Cimento} {\bf #1} (19#2) #3}
 \newcommand{\ijmp}[3]{{\sl Int. J. Mod. Phys.} {\bf #1} (19#2) #3}
 \newcommand{\ptp}[3]{{\sl Prog. Theo. Phys.} {\bf #1} (19#2) #3}
 \newcommand{\sjnp}[3]{{\sl Sov. J. Nucl. Phys.} {\bf #1} (19#2) #3}
 \newcommand{\cpc}[3]{{\sl Comp. Phys. Commun.} {\bf #1} (19#2) #3}
 \newcommand{\mpl}[3]{{\sl Mod. Phys. Lett.} {\bf #1} (19#2) #3}
 \newcommand{\cmp}[3]{{\sl Commun. Math. Phys.} {\bf #1} (19#2) #3}
 \newcommand{\jmp}[3]{{\sl J. Math. Phys.} {\bf #1} (19#2) #3}
 \newcommand{\nim}[3]{{\sl Nucl. Instr. Meth.} {\bf #1} (19#2) #3}
 \newcommand{\el}[3]{{\sl Europhysics Letters} {\bf #1} (19#2) #3}
 \newcommand{\ap}[3]{{\sl Ann. of Phys.} {\bf #1} (19#2) #3}
 \newcommand{\jetp}[3]{{\sl JETP} {\bf #1} (19#2) #3}
 \newcommand{\acpp}[3]{{\sl Acta Physica Polonica} {\bf #1} (19#2) #3}
 \newcommand{\vj}[4]{{\sl #1~}{\bf #2} (19#3) #4}
 \newcommand{\ej}[3]{{\bf #1} (19#2) #3}
 \newcommand{\vjs}[2]{{\sl #1~}{\bf #2}}
\def\ch#1{\smallskip\item[]\hskip-\labelwidth {\bf\boldmath#1}
                                  \nopagebreak}
\def\comment#1{#1\\*}
\def\ch#1{\relax}
\def\comment#1{\relax}

}
\end{document}